\documentclass[sigconf,nonacm]{acmart}

\AtBeginDocument{%
  }
\setcopyright{acmcopyright}
\copyrightyear{2026}
\acmYear{2026}
\acmDOI{XXXXXXX.XXXXXXX}
\acmConference[Conference acronym 'XX]{Make sure to enter the correct
  conference title from your rights confirmation email}{June 03--05,
  2026}{Woodstock, NY}
\acmPrice{15.00}
\acmISBN{978-1-4503-XXXX-X/26/06}

\usepackage{multirow}
\usepackage{algorithm}
\usepackage{algpseudocode}
\usepackage[table]{xcolor}

\definecolor{gaincolor}{RGB}{0,125,70}
\definecolor{losscolor}{RGB}{200,35,35}
\definecolor{oursbg}{RGB}{228,238,250}
\newcommand{\gain}[1]{\textcolor{gaincolor}{#1}}
\newcommand{\loss}[1]{\textcolor{losscolor}{#1}}
\newcommand{\ours}{\cellcolor{oursbg}}

\newcommand{\NBAllRepN}{14}
\newcommand{\NBAllRepPos}{11}

\newcommand{\NBDeepRepN}{8}

\newcommand{\NBDeepSignP}{0.008}

\newcommand{\NImpSpearman}{0.94}

\newcommand{\NLogRegBDelta}{$-$0.10}

\newcommand{\NLogRegNearHalf}{34}

\newcommand{\NMaxBase}{66.27}
\newcommand{\NMaxBaseName}{random forest}
\newcommand{\NMinBase}{63.74}
\newcommand{\NMinBaseName}{ResNet}

\newcommand{\NRFBDelta}{$-$0.50}

\newcommand{\NRFNearHalf}{23}

\newcommand{\NResNetAucComb}{0.66}
\newcommand{\NResNetAucCombGain}{+0.016}
\newcommand{\NResNetAucCombGainHi}{+0.033}
\newcommand{\NResNetAucCombGainLo}{+0.000}
\newcommand{\NResNetAucConf}{0.64}
\newcommand{\NResNetAucDir}{0.62}

\newcommand{\NResNetBConfirm}{+0.40, +0.20}
\newcommand{\NResNetBDelta}{+0.45}

\newcommand{\NResNetBEceRef}{9.0}

\newcommand{\NResNetBP}{0.68}
\newcommand{\NResNetBRef}{64.19}

\newcommand{\NResNetBankTV}{+0.45}
\newcommand{\NResNetBase}{63.74}

\newcommand{\NResNetDBrier}{$-$5.1}

\newcommand{\NResNetDNll}{$-$16.2}
\newcommand{\NResNetDefHalfComb}{74.67}
\newcommand{\NResNetDefHalfConf}{73.24}
\newcommand{\NResNetDefSeventyComb}{70.33}

\newcommand{\NResNetDelta}{+0.50}
\newcommand{\NResNetEceBase}{11.4}

\newcommand{\NResNetFeat}{+0.20}

\newcommand{\NResNetGatedBase}{60.9}
\newcommand{\NResNetGatedFrac}{31.5}
\newcommand{\NResNetGatedRef}{62.4}
\newcommand{\NResNetHard}{$-$0.35}
\newcommand{\NResNetHighChanged}{0.1}
\newcommand{\NResNetLowBase}{53.0}

\newcommand{\NResNetLowRef}{54.2}

\newcommand{\NResNetMcGate}{+0.94}

\newcommand{\NResNetNearHalf}{10}
\newcommand{\NResNetNoGate}{$-$0.20}

\newcommand{\NResNetRepPos}{3}

\newcommand{\NResNetThresh}{+0.45}
\newcommand{\NResNetUngatedAcc}{65.0}

\newcommand{\NSVMBDelta}{+1.39}

\newcommand{\NTabMAucComb}{0.66}
\newcommand{\NTabMAucCombGain}{+0.011}
\newcommand{\NTabMAucCombGainHi}{+0.028}
\newcommand{\NTabMAucCombGainLo}{-0.007}
\newcommand{\NTabMAucConf}{0.65}
\newcommand{\NTabMAucDir}{0.61}

\newcommand{\NTabMAucRis}{0.51}
\newcommand{\NTabMAucRos}{0.53}

\newcommand{\NTabMBConfirm}{+1.98, +0.99}
\newcommand{\NTabMBDelta}{+1.04}

\newcommand{\NTabMBEceRef}{6.2}

\newcommand{\NTabMBMaxChanged}{8.3}
\newcommand{\NTabMBP}{0.38}
\newcommand{\NTabMBRef}{65.28}

\newcommand{\NTabMBankTV}{+1.04}
\newcommand{\NTabMBase}{64.24}

\newcommand{\NTabMDBrier}{$-$3.6}

\newcommand{\NTabMDNll}{$-$10.4}
\newcommand{\NTabMDefHalfComb}{75.00}
\newcommand{\NTabMDefHalfConf}{75.34}
\newcommand{\NTabMDefSeventyComb}{70.58}

\newcommand{\NTabMDelta}{+0.05}
\newcommand{\NTabMEceBase}{7.6}

\newcommand{\NTabMFeat}{+0.64}

\newcommand{\NTabMGatedBase}{60.4}
\newcommand{\NTabMGatedFrac}{35.3}
\newcommand{\NTabMGatedRef}{63.3}
\newcommand{\NTabMHard}{$-$0.45}
\newcommand{\NTabMHighChanged}{0.1}
\newcommand{\NTabMLowBase}{51.5}

\newcommand{\NTabMLowRef}{54.0}

\newcommand{\NTabMMcGate}{+0.60}

\newcommand{\NTabMNearHalf}{12}
\newcommand{\NTabMNoGate}{$-$0.74}

\newcommand{\NTabMPrecFeatBest}{57.6}
\newcommand{\NTabMPrecShapEuc}{58.9}

\newcommand{\NTabMRepN}{4}
\newcommand{\NTabMRepPos}{2}
\newcommand{\NTabMRetFeatBest}{61.3}
\newcommand{\NTabMRetShapEuc}{62.2}

\newcommand{\NTabMThresh}{$-$0.30}
\newcommand{\NTabMUngatedAcc}{66.3}

\providecommand{\NBAllRepN}{??}
\providecommand{\NBAllRepPos}{??}
\providecommand{\NBDeepRepN}{??}
\providecommand{\NBDeepSignP}{??}
\providecommand{\NImpSpearman}{??}
\providecommand{\NLogRegBDelta}{??}
\providecommand{\NLogRegNearHalf}{??}
\providecommand{\NMaxBase}{??}
\providecommand{\NMaxBaseName}{??}
\providecommand{\NMinBase}{??}
\providecommand{\NMinBaseName}{??}
\providecommand{\NResNetAucComb}{??}
\providecommand{\NResNetAucCombGain}{??}
\providecommand{\NResNetAucCombGainHi}{??}
\providecommand{\NResNetAucCombGainLo}{??}
\providecommand{\NResNetAucConf}{??}
\providecommand{\NResNetAucDir}{??}
\providecommand{\NResNetBankTV}{??}
\providecommand{\NResNetBase}{??}
\providecommand{\NResNetBConfirm}{??}
\providecommand{\NResNetBDelta}{??}
\providecommand{\NResNetBEceRef}{??}
\providecommand{\NResNetBP}{??}
\providecommand{\NResNetBRef}{??}
\providecommand{\NResNetDBrier}{??}
\providecommand{\NResNetDefHalfComb}{??}
\providecommand{\NResNetDefHalfConf}{??}
\providecommand{\NResNetDefSeventyComb}{??}
\providecommand{\NResNetDelta}{??}
\providecommand{\NResNetDNll}{??}
\providecommand{\NResNetEceBase}{??}
\providecommand{\NResNetFeat}{??}
\providecommand{\NResNetGatedBase}{??}
\providecommand{\NResNetGatedFrac}{??}
\providecommand{\NResNetGatedRef}{??}
\providecommand{\NResNetHard}{??}
\providecommand{\NResNetHighChanged}{??}
\providecommand{\NResNetLowBase}{??}
\providecommand{\NResNetLowRef}{??}
\providecommand{\NResNetMcGate}{??}
\providecommand{\NResNetNearHalf}{??}
\providecommand{\NResNetNoGate}{??}
\providecommand{\NResNetRepPos}{??}
\providecommand{\NResNetThresh}{??}
\providecommand{\NResNetUngatedAcc}{??}
\providecommand{\NRFBDelta}{??}
\providecommand{\NRFNearHalf}{??}
\providecommand{\NSVMBDelta}{??}
\providecommand{\NTabMAucComb}{??}
\providecommand{\NTabMAucCombGain}{??}
\providecommand{\NTabMAucCombGainHi}{??}
\providecommand{\NTabMAucCombGainLo}{??}
\providecommand{\NTabMAucConf}{??}
\providecommand{\NTabMAucDir}{??}
\providecommand{\NTabMAucRis}{??}
\providecommand{\NTabMAucRos}{??}
\providecommand{\NTabMBankTV}{??}
\providecommand{\NTabMBase}{??}
\providecommand{\NTabMBConfirm}{??}
\providecommand{\NTabMBDelta}{??}
\providecommand{\NTabMBEceRef}{??}
\providecommand{\NTabMBMaxChanged}{??}
\providecommand{\NTabMBP}{??}
\providecommand{\NTabMBRef}{??}
\providecommand{\NTabMDBrier}{??}
\providecommand{\NTabMDefHalfComb}{??}
\providecommand{\NTabMDefHalfConf}{??}
\providecommand{\NTabMDefSeventyComb}{??}
\providecommand{\NTabMDelta}{??}
\providecommand{\NTabMDNll}{??}
\providecommand{\NTabMEceBase}{??}
\providecommand{\NTabMFeat}{??}
\providecommand{\NTabMGatedBase}{??}
\providecommand{\NTabMGatedFrac}{??}
\providecommand{\NTabMGatedRef}{??}
\providecommand{\NTabMHard}{??}
\providecommand{\NTabMHighChanged}{??}
\providecommand{\NTabMLowBase}{??}

\providecommand{\NTabMLowRef}{??}
\providecommand{\NTabMMcGate}{??}
\providecommand{\NTabMNearHalf}{??}
\providecommand{\NTabMNoGate}{??}
\providecommand{\NTabMPrecFeatBest}{??}
\providecommand{\NTabMPrecShapEuc}{??}
\providecommand{\NTabMRepN}{??}
\providecommand{\NTabMRepPos}{??}
\providecommand{\NTabMRetFeatBest}{??}
\providecommand{\NTabMRetShapEuc}{??}
\providecommand{\NTabMThresh}{??}
\providecommand{\NTabMUngatedAcc}{??}

\begin{document}

\title{Uncertainty-Gated SHAP--Dirichlet Refinement for MRI-Based Depression Classification}

\author{Phuc Truong Loc Nguyen}
\authornote{Both authors contributed equally to this research.}
\affiliation{%
  \institution{FAU Erlangen-N\"urnberg}
  \city{Erlangen}
  \country{Germany}}
\email{loc.pt.nguyen@fau.de}

\author{Nguyen Truong Toan To}
\authornotemark[1]
\affiliation{%
  \institution{FAU Erlangen-N\"urnberg}
  \city{Erlangen}
  \country{Germany}}
\email{toan.nt.to@fau.de}

\author{Truong Thanh Hung Nguyen}
\authornote{Corresponding author.}
\affiliation{%
  \institution{University of New Brunswick}
  \city{Fredericton}
  \country{Canada}}
\email{hung.ntt@unb.ca}

\author{Hoang-Loc Cao}
\affiliation{%
  \institution{University of New Brunswick}
  \city{Fredericton}
  \country{Canada}}
\email{chloc22@clc.fitus.edu.vn}

\author{Hung Cao}
\affiliation{%
  \institution{University of New Brunswick}
  \city{Fredericton}
  \country{Canada}}
\email{hung.cao@unb.ca}

\renewcommand{\shortauthors}{Nguyen et al.}

\begin{abstract}
Structural MRI-based classification of major depressive disorder (MDD) remains challenging due to subtle, heterogeneous anatomical changes. We propose uncertainty-gated SHAP-Dirichlet refinement, a post-hoc reliability layer that improves predictions without retraining the classifier. The method retrieves similar labeled SHAP explanations, constructs Dirichlet class evidence, and selectively refines uncertain predictions through probabilistic fusion, while allowing unreliable cases to be deferred for expert review. We evaluate five classifiers on 504 SRPBS subjects (252 controls and 252 MDD patients) using 159 anatomical features and repeated nested cross-validation. Refinement improved accuracy across all \NBDeepRepN{} repetitions for TabM and ResNet, yielding mean gains of \NTabMBDelta{} and \NResNetBDelta{} percentage points (sign test $p=\NBDeepSignP$), while modifying at most \NTabMBMaxChanged\% of predictions. Gains were concentrated among low-confidence cases, with significant improvements in calibration. The method also benefited an RBF support vector machine but not logistic regression or random forests. Ablations confirmed the importance of uncertainty gating and soft fusion. Dirichlet evidence identified baseline errors (AUROC \NResNetAucDir{}), while deferring 50\% of cases increased retained-case accuracy to approximately 75\%. These findings support selective explanation-guided refinement as a practical approach to improving prediction reliability and facilitating expert triage in MDD classification.
\end{abstract}




\maketitle

\section{Introduction}

Major depressive disorder (MDD) is a heterogeneous psychiatric disorder whose diagnosis is based on clinical symptoms rather than a single biological marker~\cite{Malhi2018}. Structural magnetic resonance imaging (MRI) provides non-invasive measurements of brain morphology, and large consortia have reported MDD-related differences in cortical thickness, surface area, and subcortical volume~\cite{Schmaal2017}. These effects are small and heterogeneous, and they are sensitive to acquisition, preprocessing, site composition, and diagnostic labeling~\cite{Woo2017,FORTIN2018104}. As a result, machine learning studies of depression report a wide range of accuracies~\cite{Kambeitz2017,Stolicyn2020}, and small samples tend to overestimate accuracy~\cite{Flint2021}. With a few hundred subjects, cross-validated accuracy also has large error bars~\cite{Varoquaux2018}, so a single split can overstate or understate the effect of any modeling choice.

For clinical decision support, average accuracy is not sufficient. A classifier may assign a confident label to a subject whose local feature pattern is unstable, weakly supported by similar cases, or inconsistent with the evidence the model uses~\cite{GHESU2021101855,Kompa2021,Begoli2019,11715640}. Uncertainty quantification methods such as Monte Carlo dropout, deep ensembles, evidential learning, calibration, and selective prediction address part of this problem~\cite{pmlr-v48-gal16,NIPS2017_9ef2ed4b,NEURIPS2018_a981f2b7,guo2017calibration,NIPS2017_4a8423d5,nguyen2026heart2mind}, but they often require retraining, architectural changes, or repeated inference. For a classifier that is already trained, a post-hoc reliability layer is attractive. Model confidence alone, however, does not indicate whether a prediction is supported by subjects whose explanations look alike.

We propose \textit{Uncertainty-gated SHAP-Dirichlet Refinement}, a post-hoc layer for a fixed probabilistic classifier. For each query subject, SHapley Additive exPlanations (SHAP)~\cite{3295222.3295230} give a local explanation vector, which is compared with a bank of labeled reference explanations. The most similar references are converted into class-level Dirichlet evidence~\cite{NEURIPS2018_a981f2b7}. An uncertainty gate, by default local explanation instability under small feature perturbations measured by relative input and output stability (RIS and ROS)~\cite{agarwal2022rethinking}, decides which predictions may change. Unstable predictions are refined by fusing the classifier probability with the explanation-derived evidence, and the least reliable subjects can be routed to expert review. The layer never changes the classifier, and every parameter is selected on validation data.

We evaluate the layer on 504 subjects of the SRPBS dataset~\cite{Tanaka2021} with five classifiers: the tabular deep learning models TabM~\cite{gorishniy2025tabm} and ResNet~\cite{3540261.3541708}, and logistic regression, an RBF support vector machine~\cite{Cortes1995}, and a random forest~\cite{Breiman2001}. Because single-split results on this task are unstable, we use repeated nested cross-validation with pre-specified statistical tests. Our contributions are as follows:

\textbf{1.} We formulate a post-hoc, reliability-aware decision layer that retains, refines, or defers the predictions of a fixed classifier by combining explanation retrieval, Dirichlet evidence, and an uncertainty gate (Sect.~\ref{sec:method}).

\textbf{2.} We describe an exact batched implementation of KernelSHAP with common random numbers, which makes explanation-stability estimates comparable across perturbations, and a fix for a batch-dependent preprocessing step in a popular tabular deep learning library (Sect.~\ref{sec:impl}).

\textbf{3.} In repeated nested cross-validation with five classifiers, the layer with a training-plus-validation reference bank improved accuracy in every TabM and ResNet repetition, concentrated its gains on low-confidence subjects, and improved calibration. Controlled ablations show that gating and soft fusion keep post-hoc changes safe, and that the size of the reference bank matters more than the choice of gate (Sects.~\ref{sec:main}--\ref{sec:calib}).

\textbf{4.} SHAP-space retrieval yields more label-consistent references than raw-feature retrieval, SHAP-Dirichlet evidence identifies baseline errors above chance and adds to model confidence, and a deferral policy built on both raises the accuracy of automatically handled subjects to about 75\% when half of the subjects are reviewed (Sects.~\ref{sec:space}--\ref{sec:review}).

\section{Related Work}
\label{sec:related}

\paragraph{Machine learning for MDD from structural MRI}
Morphometric features from FreeSurfer or FastSurfer~\cite{FISCHL2012774,HENSCHEL2020117012}, typically summarized on the Desikan-Killiany atlas~\cite{Desikan2006}, are a common input for MDD classification. Group-level analyses of large consortia show consistent but small cortical and subcortical differences~\cite{Schmaal2017}. A meta-analysis of multivariate pattern recognition studies found heterogeneous accuracies across studies and modalities~\cite{Kambeitz2017}. In community cohorts, a cross-validated accuracy of 75\% in a small sample dropped to 53.7\% in a larger independent sample~\cite{Stolicyn2020}, and a systematic study showed that small samples overestimate performance~\cite{Flint2021}. Cross-validation in small samples has large error bars~\cite{Varoquaux2018}, and methodological pitfalls such as test-set reuse remain common in medical imaging~\cite{Varoquaux2022}. These findings motivate our use of repeated nested cross-validation and our focus on reliability rather than on a single accuracy number.

\paragraph{Uncertainty, calibration, and selective prediction}
Monte Carlo dropout~\cite{pmlr-v48-gal16}, deep ensembles~\cite{NIPS2017_9ef2ed4b}, and evidential deep learning~\cite{NEURIPS2018_a981f2b7} estimate predictive uncertainty but change how the model is trained or queried. Calibration methods~\cite{guo2017calibration} and calibration metrics such as the expected calibration error (ECE)~\cite{Naeini2015} and the Brier score~\cite{Brier1950} assess whether probabilities match observed frequencies. Selective prediction abstains on uncertain inputs~\cite{Chow1970,NIPS2017_4a8423d5}, with the maximum class probability as a strong baseline~\cite{Hendrycks2017}, and learning-to-defer methods train a classifier together with a deferral policy~\cite{Mozannar2020}. In medical imaging, referring uncertain cases to experts improves the accuracy of the remaining automatic decisions~\cite{Leibig2017,GHESU2021101855}, and communicating uncertainty is considered essential for clinical use~\cite{Kompa2021,Begoli2019}. Our layer is post hoc, keeps the classifier fixed, and combines deferral with a correction step for unstable predictions.

\paragraph{Neighbors and explanations as reliability signals}
Nearest-neighbor agreement is a known reliability signal. The trust score compares the distance to the predicted class with the distance to other classes, and deep $k$-nearest neighbors measure label agreement in learned representations~\cite{Papernot2018}. We instead retrieve neighbors in the explanation space, where distances reflect how the model uses each feature, and convert them into Dirichlet evidence. SHAP values~\cite{3295222.3295230,Shapley1953} are commonly estimated with KernelSHAP~\cite{3295222.3295230,Covert2021}. The stability of explanations under small input changes has been studied as a robustness property~\cite{AlvarezMelis2018,agarwal2022rethinking}, and toolkits such as Quantus implement RIS and ROS~\cite{Hedstrom2023}. We use RIS and ROS not to evaluate explanation methods, but to gate which predictions a post-hoc layer may change.

\paragraph{Tabular classifiers}
Deep tabular models such as ResNet-style networks~\cite{3540261.3541708} and the parameter-efficient ensemble TabM~\cite{gorishniy2025tabm} are strong general-purpose baselines, while tree-based models remain competitive with deep models on typical tabular data~\cite{Grinsztajn2022}. We therefore evaluate the layer on both neural and classical classifiers~\cite{Pedregosa2011}.

\section{Method}
\label{sec:method}

\begin{figure*}[t]
\centering
\includegraphics[width=0.7\textwidth]{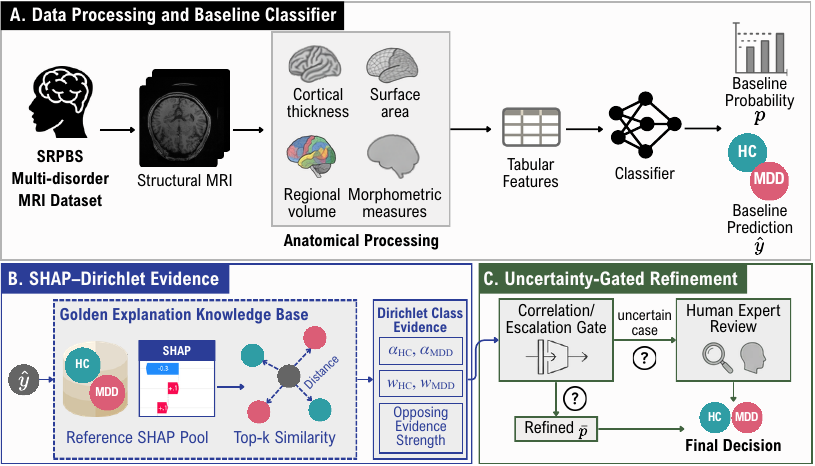}
\caption{Uncertainty-gated SHAP-Dirichlet refinement. (A) Structural MRI scans are processed into morphometric features, and a fixed classifier produces the baseline probability and label. (B) The query SHAP explanation is compared with a reference SHAP bank to retrieve explanation-similar subjects and estimate Dirichlet class evidence. (C) Local explanation stability decides whether the prediction is retained or refined, and unreliable subjects can be routed to expert review.}
\label{fig:framework}
\end{figure*}

Fig.~\ref{fig:framework} summarizes the framework. It is applied after training and leaves the classifier unchanged. Its three components are the baseline classifier, SHAP-Dirichlet evidence, and uncertainty-gated refinement with an optional deferral step.

\subsection{Baseline Classifier}

Given a structural MRI scan $I_i$, a morphometric feature extractor $F(\cdot)$ maps the image to a tabular representation $x_i=F(I_i)\in\mathbb{R}^{d}$. Here, $F(\cdot)$ is FreeSurfer/FastSurfer, which returns regional cortical and subcortical measurements~\cite{HENSCHEL2020117012,FISCHL2012774}. A classifier $f_\theta$ is trained on these features. For a query subject $q$, it outputs the MDD probability $p_q=f_\theta(x_q)$, and the baseline label is $\hat{y}_q=\mathbf{1}(p_q\geq0.5)$. After training, $\theta$ is frozen. The refinement stage only uses the model output and its local explanation, so any classifier that outputs a probability can be used.

\subsection{SHAP-Dirichlet Evidence}

SHAP~\cite{3295222.3295230} gives a local attribution vector $\phi_q=[\phi_{q1},\ldots,\phi_{qd}]$ of the MDD probability, where each entry measures the contribution of one feature. Because the explained output is fixed, a positive entry pushes the prediction toward MDD and a negative entry toward HC for every subject. The query explanation is compared with a reference bank $\mathcal{R}$ of explanations of labeled subjects, and a subject never retrieves itself. We consider two banks: the validation subjects only, and the training and validation subjects together, which are all labeled subjects available when the layer is deployed. The top-$K$ most similar references are
\begin{equation}
\label{eq:neighborhood}
\mathcal{N}_K(q)=\operatorname{TopK}_{r\in\mathcal{R}}\big(\operatorname{sim}(\phi_q,\phi_r)\big),
\end{equation}
where $\operatorname{sim}(\cdot,\cdot)$ is a cosine, Pearson, or negative Euclidean similarity. The retrieved references are converted into class-level evidence
\begin{equation}
\label{eq:class_evidence}
e_c(q)=\sum_{r\in\mathcal{N}_K(q)} w_{qr}\mathbf{1}(y_r=c),\qquad w_{qr}=\exp\!\left(-\frac{d(\phi_q,\phi_r)}{\tau_q}\right),
\end{equation}
where $d(\cdot,\cdot)$ is the distance associated with the similarity and $\tau_q$ is the median distance within $\mathcal{N}_K(q)$. Adding one unit of prior evidence to each class gives the Dirichlet-derived class probability
\begin{equation}
\label{eq:dirichlet_probability}
p_c^{\mathrm{Dir}}(q)=\frac{e_c(q)+1}{\sum_{c'\in\{0,1\}}\big(e_{c'}(q)+1\big)}.
\end{equation}
The Dirichlet probability is not a second classifier. It estimates how strongly the query explanation agrees with reference subjects of each class, and the prior keeps it close to $0.5$ when the evidence is weak. The explanation-supported class is $\tilde{y}_q=\arg\max_{c}p_c^{\mathrm{Dir}}(q)$, and the support for the class opposite to the baseline label is $E_q^{\mathrm{opp}}=p^{\mathrm{Dir}}_{1-\hat{y}_q}(q)$.

\subsection{Uncertainty-Gated Refinement}

\paragraph{Local explanation instability} We estimate how much the explanation of a query changes under small feature perturbations, using relative input stability (RIS) and relative output stability (ROS)~\cite{agarwal2022rethinking}. For each query, we draw perturbed samples $x_q^{(m)}=x_q+\delta_q^{(m)}$, $m=1,\ldots,M$, and keep only those that preserve the predicted label. For each kept perturbation, we compute the explanation $\phi_q^{(m)}$ and the model logit $h_\theta(x_q^{(m)})$, and define
\begin{align}
\mathrm{RIS}(q)&=\max_m\frac{\big\|(\phi_q-\phi_q^{(m)})/(\phi_q+\epsilon)\big\|_p}{\max\big(\|(x_q-x_q^{(m)})/(x_q+\epsilon)\|_p,\ \epsilon_{\min}\big)},\label{eq:ris}\\
\mathrm{ROS}(q)&=\max_m\frac{\big\|(\phi_q-\phi_q^{(m)})/(\phi_q+\epsilon)\big\|_p}{\max\big(\|h_\theta(x_q)-h_\theta(x_q^{(m)})\|_p,\ \epsilon_{\min}\big)}.\label{eq:ros}
\end{align}
Larger values indicate lower local reliability, and $\epsilon$ and $\epsilon_{\min}$ prevent division by zero. If no perturbation preserves the predicted label, the query is treated as maximally unstable.

\paragraph{Gate} Let $u\in\{\mathrm{RIS},\mathrm{ROS}\}$ be the selected instability measure and $\eta_q^{(u)}=u(q)$. Other uncertainty scores, such as Monte Carlo dropout variance, can take the place of $\eta^{(u)}_q$, and we compare them in Sect.~\ref{sec:ablation}. For a validation percentile $\pi$, the threshold $\tau_{u,\pi}$ is the $\pi$-th percentile of $\eta^{(u)}$ on the validation set, and the gate is
\begin{equation}
\label{eq:gate}
g_q=\mathbf{1}\big(\eta_q^{(u)}\geq\tau_{u,\pi}\big).
\end{equation}
A stable subject ($g_q=0$) keeps its baseline prediction. An unstable subject ($g_q=1$) is passed to the correction rule.

\paragraph{Correction} Soft correction blends the baseline probability with the Dirichlet-derived MDD probability,
\begin{equation}
\label{eq:soft}
p^{\mathrm{soft}}_q=\begin{cases}\lambda p_q+(1-\lambda)\,p^{\mathrm{Dir}}_1(q), & g_q=1,\\ p_q, & g_q=0,\end{cases}
\end{equation}
and the refined label is $y^{\mathrm{soft}}_q=\mathbf{1}(p^{\mathrm{soft}}_q\geq0.5)$. As an alternative, hard correction flips the label of a gated subject when $\tilde{y}_q\neq\hat{y}_q$ and $E_q^{\mathrm{opp}}\geq\tau_E$.

\paragraph{Validation-based selection} The instability measure $u$, the similarity, $\pi$, $\lambda$, and $\tau_E$ are selected on the validation set by accuracy. Ties are broken toward the rule that changes the fewest validation decisions, and if no rule improves validation accuracy, the baseline is kept for every subject. Algorithm~\ref{alg:refine} summarizes the procedure.

\begin{algorithm}[t]
\caption{Uncertainty-gated SHAP-Dirichlet refinement}
\label{alg:refine}
\small
\begin{algorithmic}[1]
\Require fixed classifier $f_\theta$, validation set $\mathcal{V}$, reference bank $\mathcal{R}\supseteq\mathcal{V}$, test subject $q$, grids $\Pi$, $\Lambda$
\State Compute $\phi_r$ and $\eta_r^{(u)}$ for all $r\in\mathcal{V}$ and $u\in\{\mathrm{RIS},\mathrm{ROS}\}$
\State For each $r\in\mathcal{V}$, compute $p^{\mathrm{Dir}}_1(r)$ from $\mathcal{R}\setminus\{r\}$ \Comment{Eqs.~\ref{eq:neighborhood}--\ref{eq:dirichlet_probability}}
\State $(u^\star,\mathrm{sim}^\star,\pi^\star,\lambda^\star)\gets$ the rule with the highest validation accuracy, ties toward fewer changes
\If{no rule improves validation accuracy} \Return $\hat{y}_q$
\EndIf
\State Compute $p_q$, $\phi_q$, $\eta^{(u^\star)}_q$, and $p^{\mathrm{Dir}}_1(q)$ with bank $\mathcal{R}$
\If{$\eta^{(u^\star)}_q\geq\tau_{u^\star,\pi^\star}$} $p_q\gets\lambda^\star p_q+(1-\lambda^\star)p^{\mathrm{Dir}}_1(q)$
\EndIf
\State \Return $\mathbf{1}(p_q\geq0.5)$ and a review score (Sect.~\ref{sec:review_method})
\end{algorithmic}
\end{algorithm}

\subsection{Expert Review}
\label{sec:review_method}

The framework also supports human-in-the-loop verification. A subject with low baseline confidence $1-|2p_q-1|$, or with high support for the opposite class $E^{\mathrm{opp}}_q$, can be routed to expert inspection instead of being decided automatically. We map both scores to their percentiles in the validation distribution, average them, and defer the highest-ranked subjects. The deferral threshold is a validation percentile, so the expected review workload is fixed before deployment.

\subsection{Implementation}
\label{sec:impl}

\paragraph{Exact batched KernelSHAP} We explain the MDD probability with KernelSHAP~\cite{3295222.3295230}, using 20 $k$-means centroids of the training split as background and 800 coalition samples. We reseed the sampler before every explanation, so a query and its perturbations share the same coalitions (common random numbers). Differences between their explanations then reflect the model rather than Monte Carlo noise of KernelSHAP, which matters because RIS and ROS divide explanation differences by small input changes. With fixed coalitions, the KernelSHAP estimate is a fixed linear map of the coalition values. We therefore take the masks and kernel weights from the reference implementation and evaluate all perturbations of a query in one batched model call. On every outer fold, the batched estimate matched the reference implementation up to floating-point precision.

\paragraph{Batch-independent preprocessing} Both neural backbones use piecewise linear encoding (PLE) of the features. The PLE transform of the library we used clamps its last bin edge to the maximum of the current batch, so the prediction for a subject depends on the other subjects in the batch. This matters for post-hoc analysis, because perturbed copies of one subject are evaluated in batches of very different composition. We removed the clamp and compiled the per-feature encoding into one vectorized GPU computation. Training inputs are unaffected, because the training maximum always exceeds the last split, and the fixed transform reproduced the reported baseline of a previous single-split experiment exactly.

\section{Experimental Setup}
\label{sec:setup}

\paragraph{Data} We used T1-weighted MRI scans from the SRPBS dataset~\cite{Tanaka2021} and formed a balanced cohort of 504 subjects, with 252 healthy controls and 252 patients with MDD. Preprocessing followed the FastSurfer/FreeSurfer-based protocol~\cite{HENSCHEL2020117012,FISCHL2012774} of Beliveau et al.~\cite{BELIVEAU2022103224}. Scans were processed automatically, followed by a quality check that excluded large segmentation failures, without manual editing. Each subject was represented by 159 features: 70 cortical thickness and 70 surface-area features (34 Desikan-Killiany regions~\cite{Desikan2006} per hemisphere plus one hemisphere average), 18 bilateral volumes of the accumbens, amygdala, caudate, cerebellar cortex, hippocampus, pallidum, putamen, thalamus, and lateral ventricles, and the estimated total intracranial volume.

\paragraph{Classifiers} TabM~\cite{gorishniy2025tabm} is an MLP with parameter-efficient implicit ensembling (32 members), and ResNet~\cite{3540261.3541708} is a residual feed-forward network. Both use PLE with 20 bins and are trained for up to 200 epochs with early stopping (patience 20) and batch size 32. Each has a grid of 72 configurations over width or depth, dropout, learning rate, and weight decay. The classical classifiers are L2-regularized logistic regression, an RBF support vector machine with Platt-scaled probabilities, and a random forest with 300 trees, each on standardized features with a small grid (Appendix~\ref{app:grids}).

\paragraph{Protocol} We used stratified 5-fold outer cross-validation, so every subject is a test subject exactly once per repetition. Within each outer fold, a 3-fold inner GridSearch selected the hyperparameters by F1-score. The selected model was refit on two thirds of the outer-training data (about 270 subjects), and the remaining third (about 135 subjects) served as the validation set for early stopping and for all refinement choices. The whole protocol was repeated with four split seeds for TabM and ResNet (20 outer folds each) and with two split seeds for the classical classifiers, whose KernelSHAP runs were the most expensive per subject. No test subject was used for training, model selection, reference construction, or rule selection.

\paragraph{Refinement settings} We used $K=5$, $M=10$ multiplicative perturbations $x_q\odot(1+u)$ with $u\sim\mathcal{U}[-0.02,0.02]^d$, the $\ell_2$ norm in RIS and ROS, and log-probabilities as the logit surrogate. The grids were $\pi\in\{50,60,70,80,90\}$, $\lambda\in\{0,0.05,\ldots,0.95\}$, and $\tau_E\in\{0.50,0.55,\ldots,0.75\}$. Unless stated otherwise, the layer uses soft fusion with the gate measure and the similarity selected on validation. We report two reference banks. The pre-specified bank contains the validation subjects (about 135). The enlarged bank adds the training subjects (about 400 in total). We noted the enlarged bank as the most promising design choice after the first two repetitions, before the remaining two repetitions were analyzed, and we report those two as a confirmation.

\paragraph{Comparisons} Each ablation changes one component and keeps every other choice validation-selected: hard instead of soft correction, gates based on no gating, a random subset of the same size, low baseline confidence, or Monte Carlo dropout variance, retrieval in the standardized raw feature space, larger reference banks, and other values of $K$. We also report decision-threshold tuning on validation data as a simple post-hoc alternative.

\paragraph{Metrics and statistics} Within a repetition, metrics are computed on the 504 pooled out-of-fold predictions. We report accuracy, balanced accuracy, macro F1-score, ROC-AUC, the Brier score~\cite{Brier1950}, the expected calibration error (ECE, 10 bins)~\cite{Naeini2015}, and the negative log-likelihood (NLL). The paired change in correctness is tested with an exact McNemar test per repetition~\cite{Dietterich1998}. Across repetitions, we use a sign test over repetition-level accuracy changes and the corrected resampled $t$-test~\cite{Nadeau2003} over all fold-level changes, which accounts for the overlap between training sets. Calibration changes are reported with 95\% paired bootstrap intervals. Error detection is measured by the AUROC of separating incorrect from correct baseline decisions, after mapping each score to its validation percentile within the fold.

\section{Results}
\label{sec:results}

\subsection{Refinement Across Classifiers}
\label{sec:main}

\begin{table*}[t]
\centering
\caption{Out-of-fold test performance of the fixed baselines and the refinement layer with the pre-specified validation reference bank and with the enlarged training+validation bank, averaged over cross-validation repetitions (each pooled over $n=504$ subjects). Accuracy, balanced accuracy (BAcc), macro F1 (MF1), and ROC-AUC are in \%, with the change in percentage points in parentheses. ECE is in \%. \gain{Green} (\loss{red}): better (worse) than the baseline of the same classifier. Reps: number of repetitions. I/R: repetitions with improved accuracy. $p$: corrected resampled $t$-test over all fold-level accuracy changes.}
\label{tab:main}
\small
\setlength{\tabcolsep}{3.6pt}
\resizebox{\textwidth}{!}{%
\begin{tabular}{lllrrrrrrcc}
\toprule
\textbf{Classifier} & \textbf{Reps} & \textbf{Model} & \textbf{Accuracy} & \textbf{BAcc} & \textbf{MF1} & \textbf{AUC} & \textbf{Brier}$\downarrow$ & \textbf{ECE}$\downarrow$ & \textbf{I/R} & $p$ \\
\midrule
\multirow{3}{*}{TabM} & \multirow{3}{*}{4} & Baseline & 64.24 & 64.24 & 64.19 & 70.92 & 0.2217 & 7.56 & & \\
 & & Layer, val.\ bank & 64.29 \gain{(+0.05)} & 64.29 \gain{(+0.05)} & 64.25 \gain{(+0.05)} & 70.84 \loss{($-$0.08)} & \gain{0.2196} & \gain{7.52} & 2/4 & 0.95 \\
 & & Layer, train+val.\ bank & 65.28 \gain{(+1.04)} & 65.28 \gain{(+1.04)} & 65.25 \gain{(+1.06)} & 71.25 \gain{(+0.33)} & \gain{0.2181} & \gain{6.24} & 4/4 & 0.38 \\
\midrule
\multirow{3}{*}{ResNet} & \multirow{3}{*}{4} & Baseline & 63.74 & 63.74 & 63.71 & 70.00 & 0.2338 & 11.44 & & \\
 & & Layer, val.\ bank & 64.24 \gain{(+0.50)} & 64.24 \gain{(+0.50)} & 64.22 \gain{(+0.51)} & 69.24 \loss{($-$0.75)} & \gain{0.2300} & \gain{8.11} & 3/4 & 0.65 \\
 & & Layer, train+val.\ bank & 64.19 \gain{(+0.45)} & 64.19 \gain{(+0.45)} & 64.18 \gain{(+0.47)} & 69.91 \loss{($-$0.09)} & \gain{0.2287} & \gain{9.03} & 4/4 & 0.68 \\
\midrule
\multirow{3}{*}{Logistic Regression} & \multirow{3}{*}{2} & Baseline & 64.29 & 64.29 & 64.28 & 69.87 & 0.2218 & 5.01 & & \\
 & & Layer, val.\ bank & 63.89 \loss{($-$0.40)} & 63.89 \loss{($-$0.40)} & 63.89 \loss{($-$0.40)} & 70.12 \gain{(+0.25)} & \gain{0.2213} & \gain{4.50} & 1/2 & 0.74 \\
 & & Layer, train+val.\ bank & 64.19 \loss{($-$0.10)} & 64.19 \loss{($-$0.10)} & 64.18 \loss{($-$0.10)} & 70.72 \gain{(+0.85)} & \gain{0.2192} & \gain{3.86} & 1/2 & 0.94 \\
\midrule
\multirow{3}{*}{RBF-SVM} & \multirow{3}{*}{2} & Baseline & 65.87 & 65.87 & 65.87 & 72.00 & 0.2127 & 4.21 & & \\
 & & Layer, val.\ bank & 66.96 \gain{(+1.09)} & 66.96 \gain{(+1.09)} & 66.95 \gain{(+1.08)} & 72.29 \gain{(+0.29)} & \gain{0.2123} & \loss{5.54} & 1/2 & 0.38 \\
 & & Layer, train+val.\ bank & 67.26 \gain{(+1.39)} & 67.26 \gain{(+1.39)} & 67.23 \gain{(+1.36)} & 72.02 \gain{(+0.02)} & \gain{0.2126} & \loss{4.25} & 2/2 & 0.39 \\
\midrule
\multirow{3}{*}{Random Forest} & \multirow{3}{*}{2} & Baseline & 66.27 & 66.27 & 66.22 & 72.46 & 0.2119 & 4.57 & & \\
 & & Layer, val.\ bank & 65.28 \loss{($-$0.99)} & 65.28 \loss{($-$0.99)} & 65.22 \loss{($-$1.01)} & 71.18 \loss{($-$1.27)} & \loss{0.2152} & \loss{6.39} & 0/2 & 0.52 \\
 & & Layer, train+val.\ bank & 65.77 \loss{($-$0.50)} & 65.77 \loss{($-$0.50)} & 65.74 \loss{($-$0.49)} & 72.03 \loss{($-$0.42)} & \loss{0.2128} & \loss{4.78} & 0/2 & 0.52 \\
\bottomrule
\end{tabular}}
\end{table*}

Table~\ref{tab:main} summarizes the main comparison. The fixed baselines reached between \NMinBase\% (\NMinBaseName) and \NMaxBase\% (\NMaxBaseName) accuracy, which is in the range reported for structural MRI in comparable samples~\cite{Kambeitz2017,Flint2021}. Fold-level accuracy varied widely, from 56\% to 73\% for TabM (Appendix~\ref{app:grids}), so single-split results on this task are not reliable.

\begin{figure*}[t]
\centering

\begin{minipage}[t]{0.28\linewidth}
    \centering
    \includegraphics[width=\linewidth]{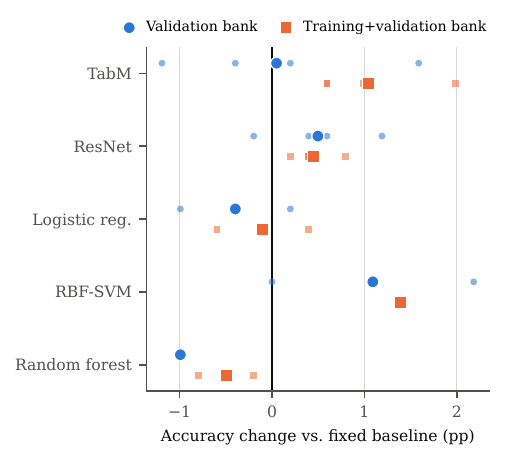}
    \caption{Accuracy change of the refined decisions relative to the fixed baseline in each cross-validation repetition (small markers) and on average (large markers), for the two reference banks.}
    \label{fig:repetitions}
\end{minipage}
\hfill
\begin{minipage}[t]{0.32\linewidth}
    \centering
    \includegraphics[width=\linewidth]{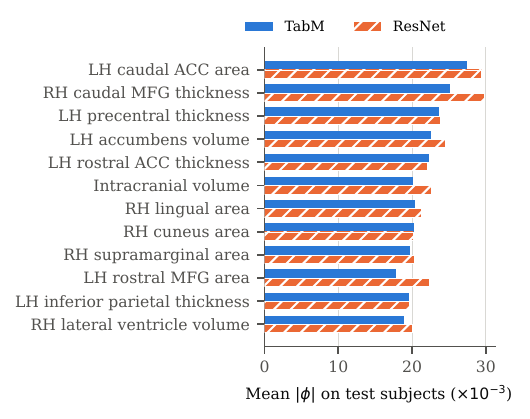}
    \caption{12 features with the largest mean absolute SHAP value on test subjects, averaged over folds and repetitions. LH/RH: left/right hemisphere. ACC: anterior cingulate cortex. MFG: middle frontal gyrus.}
    \label{fig:importance}
\end{minipage}
\hfill
\begin{minipage}[t]{0.36\linewidth}
    \centering
    \includegraphics[width=\linewidth]{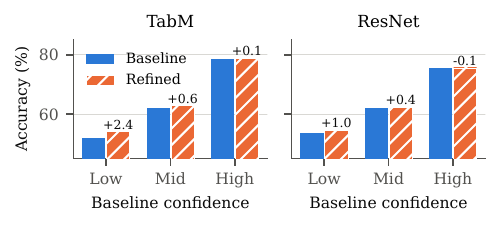}
    \caption{Accuracy by tercile of baseline confidence $|2p_q-1|$, before and after refinement with the enlarged bank, pooled over repetitions. Numbers give the change in percentage points.}
    \label{fig:terciles}
\end{minipage}

\end{figure*}

With the pre-specified validation bank, the layer changed accuracy by \NTabMDelta{} points for TabM and \NResNetDelta{} points for ResNet, with gains in \NTabMRepPos{} and \NResNetRepPos{} of \NTabMRepN{} repetitions. With about 135 references, the Dirichlet evidence is noisy, and the first two repetitions suggested that a larger bank helps (Table~\ref{tab:ablation}). With the enlarged training+validation bank, accuracy increased in all \NBDeepRepN{} neural backbone repetitions (sign test $p=\NBDeepSignP$), from \NTabMBase\% to \NTabMBRef\% for TabM and from \NResNetBase\% to \NResNetBRef\% for ResNet (Fig.~\ref{fig:repetitions}). The two confirmation repetitions gave \NTabMBConfirm{} points for TabM and \NResNetBConfirm{} points for ResNet. Balanced accuracy and macro F1 moved by the same amount, and ROC-AUC changed little because soft fusion only reorders the scores of gated subjects. The layer changed few decisions, at most \NTabMBMaxChanged\% of the subjects in any repetition. Fold-level changes remain noisy, so the corrected resampled $t$-test was not significant for either backbone ($p=\NTabMBP$ and $p=\NResNetBP$), and the gains should be read as small but consistent.

For the classical classifiers, the effect depended on the model. With the enlarged bank, refinement changed the accuracy of the RBF-SVM by \NSVMBDelta{} points (improved in both repetitions), of logistic regression by \NLogRegBDelta{} points, and of the random forest by \NRFBDelta{} points. The two classifiers that did not benefit had the most compressed probabilities: \NLogRegNearHalf\% (logistic regression) and \NRFNearHalf\% (random forest) of the test subjects had $|2p_q-1|<0.1$, compared with \NTabMNearHalf\% and \NResNetNearHalf\% for TabM and ResNet. Near $0.5$, a small shift toward the Dirichlet estimate flips decisions in both directions. Over all five classifiers, accuracy increased in \NBAllRepPos{} of \NBAllRepN{} repetitions.

\subsection{Layer Acts Localization}
\label{sec:where}

We now analyze the layer with the enlarged bank. The gate passed \NTabMGatedFrac\% (TabM) and \NResNetGatedFrac\% (ResNet) of the test subjects to the correction rule, and it selected harder subjects. Their baseline accuracy was \NTabMGatedBase\% for TabM and \NResNetGatedBase\% for ResNet, compared with \NTabMUngatedAcc\% and \NResNetUngatedAcc\% for the subjects that were not gated. Among gated subjects, refinement increased accuracy to \NTabMGatedRef\% and \NResNetGatedRef\%. Fig.~\ref{fig:terciles} splits the test subjects by baseline confidence. The gains are concentrated in the least confident third, where accuracy increased from \NTabMLowBase\% to \NTabMLowRef\% for TabM and from \NResNetLowBase\% to \NResNetLowRef\% for ResNet, while only \NTabMHighChanged\% (TabM) and \NResNetHighChanged\% (ResNet) of the most confident decisions were changed. This is the intended behavior of a post-hoc layer: it leaves confident predictions alone and uses the explanation evidence where the classifier is close to the decision boundary.

\subsection{Ablation Study}
\label{sec:ablation}

\begin{table}[t]
\centering
\caption{Ablations for the neural backbones. Each row changes one component of the pre-specified layer (soft fusion, validated RIS/ROS gate, validation bank, $K=5$), and all other choices are still selected on validation data. $\Delta$: mean accuracy change (percentage points) over repetitions, \gain{green} for a gain and \loss{red} for a loss. I/R: repetitions with improved accuracy. Shaded: the pre-specified layer.}
\label{tab:ablation}
\small
\setlength{\tabcolsep}{2.6pt}
\resizebox{\columnwidth}{!}{%
\begin{tabular}{llrrrr}
\toprule
& & \multicolumn{2}{c}{\textbf{TabM}} & \multicolumn{2}{c}{\textbf{ResNet}} \\
\textbf{Component} & \textbf{Variant} & $\Delta$ & I/R & $\Delta$ & I/R \\
\midrule
\multirow{3}{*}{Correction} & Hard flip & \loss{$-$0.45} & 1/4 & \loss{$-$0.35} & 0/4 \\
 & Hard or soft (validated) & \gain{+0.45} & 2/4 & \gain{+0.40} & 3/4 \\
 & \ours Soft fusion (ours) & \ours \gain{+0.05} & \ours 2/4 & \ours \gain{+0.50} & \ours 3/4 \\
\midrule
\multirow{7}{*}{Gate} & None & \loss{$-$0.74} & 1/4 & \loss{$-$0.20} & 1/4 \\
 & Random & \loss{$-$0.13} & 1/4 & \gain{+0.11} & 1/4 \\
 & Low confidence & \gain{+0.25} & 2/4 & \gain{+0.40} & 1/4 \\
 & MC-dropout variance & \gain{+0.60} & 2/4 & \gain{+0.94} & 4/4 \\
 & RIS only & \loss{$-$0.15} & 1/4 & \gain{+0.69} & 4/4 \\
 & ROS only & \gain{+0.69} & 3/4 & \gain{+0.20} & 2/4 \\
 & \ours RIS/ROS, validated (ours) & \ours \gain{+0.05} & \ours 2/4 & \ours \gain{+0.50} & \ours 3/4 \\
\midrule
\multirow{4}{*}{Evidence} & Raw-feature retrieval & \gain{+0.64} & 3/4 & \gain{+0.20} & 2/4 \\
 & SHAP, training bank & \gain{+0.35} & 3/4 & \gain{+0.35} & 4/4 \\
 & SHAP, training+validation bank & \gain{+1.04} & 4/4 & \gain{+0.45} & 4/4 \\
 & \ours SHAP, validation bank (ours) & \ours \gain{+0.05} & \ours 2/4 & \ours \gain{+0.50} & \ours 3/4 \\
\midrule
\multirow{5}{*}{Top-$K$} & $K=3$ & \gain{+0.20} & 3/4 & \loss{$-$0.64} & 0/4 \\
 & \ours $K=5$ (ours) & \ours \gain{+0.05} & \ours 2/4 & \ours \gain{+0.50} & \ours 3/4 \\
 & $K=7$ & \loss{$-$0.05} & 2/4 & \loss{$-$0.05} & 2/4 \\
 & $K=10$ & \gain{+0.15} & 2/4 & \loss{$-$0.25} & 1/4 \\
 & $K=15$ & \gain{+0.30} & 2/4 & \loss{$-$0.30} & 1/4 \\
\bottomrule
\end{tabular}}
\end{table}

Table~\ref{tab:ablation} changes one component of the pre-specified layer at a time. Hard flipping lowered accuracy for both backbones (\NTabMHard{} and \NResNetHard{} points), while soft fusion did not. Without a gate, soft fusion acted on every subject and changed accuracy by \NTabMNoGate{} and \NResNetNoGate{} points, because it also altered many correct decisions. Every gate was better than no gate. Monte Carlo dropout variance gave the largest gain for ResNet (\NResNetMcGate{} points) and a gain of \NTabMMcGate{} points for TabM, and the explanation-stability gates were competitive. The choice of the uncertainty measure therefore mattered less than having a gate. The reference bank was the most influential choice. Adding the training subjects gave \NTabMBankTV{} and \NResNetBankTV{} points and improved accuracy in every repetition, while retrieval in the raw feature space gave \NTabMFeat{} and \NResNetFeat{} points. The default $K=5$ was the best value for ResNet, and no value of $K$ changed TabM accuracy by more than 0.3 points. Decision-threshold tuning alone gave \NTabMThresh{} and \NResNetThresh{} points.

\subsection{Calibration}
\label{sec:calib}

\begin{table}[t]
\centering
\caption{Change in calibration after refinement with the enlarged bank (refined minus baseline, lower is better), averaged over repetitions, with 95\% paired bootstrap intervals. Brier and NLL are $\times10^{3}$, ECE is in percentage points. \gain{Green} with $^{*}$: improvement whose interval excludes zero.}
\label{tab:calib}
\small
\setlength{\tabcolsep}{2.4pt}
\resizebox{\columnwidth}{!}{%
\begin{tabular}{lccc}
\toprule
\textbf{Classifier} & $\Delta$\textbf{Brier} & $\Delta$\textbf{ECE} & $\Delta$\textbf{NLL} \\
\midrule
TabM & \gain{$-$3.6$^{*}$ [$-$6.2, $-$1.1]} & $-$1.3 [$-$2.5, 0.2] & \gain{$-$10.4$^{*}$ [$-$18.2, $-$3.5]} \\
ResNet & \gain{$-$5.1$^{*}$ [$-$8.5, $-$1.9]} & \gain{$-$2.4$^{*}$ [$-$3.2, $-$0.7]} & \gain{$-$16.2$^{*}$ [$-$25.8, $-$7.4]} \\
Logistic regression & \gain{$-$2.6$^{*}$ [$-$5.0, $-$0.2]} & $-$1.1 [$-$3.1, 1.4] & $-$8.0 [$-$19.0, 0.3] \\
RBF-SVM & $-$0.2 [$-$2.6, 2.4] & +0.0 [$-$2.2, 1.8] & +0.1 [$-$5.7, 5.8] \\
Random forest & +0.9 [$-$0.4, 2.3] & +0.2 [$-$1.6, 1.9] & +1.9 [$-$1.0, 4.9] \\
\bottomrule
\end{tabular}}
\end{table}

\begin{figure}[ht]
\centering
\includegraphics[width=.9\linewidth]{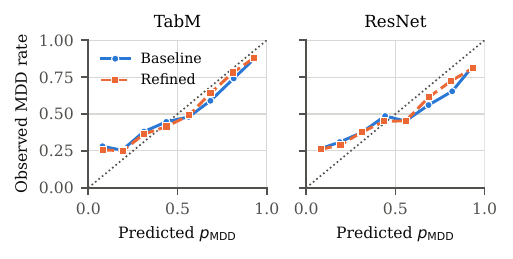}
\caption{Reliability diagrams of the baseline and refined MDD probabilities (enlarged bank), pooled over repetitions. The dotted line marks perfect calibration.}
\label{fig:calibration}
\end{figure}

Refinement also improved the probabilities themselves (Table~\ref{tab:calib}). For both neural backbones, the Brier score and the NLL decreased, and the 95\% intervals exclude zero (TabM \NTabMDBrier{} and \NTabMDNll{}, ResNet \NResNetDBrier{} and \NResNetDNll{}, both $\times10^{-3}$). The ECE of ResNet dropped from \NResNetEceBase\% to \NResNetBEceRef\%, and that of TabM from \NTabMEceBase\% to \NTabMBEceRef\%. Logistic regression also gained in Brier score, while the SVM and the random forest were unchanged. Fig.~\ref{fig:calibration} shows the reason. ResNet was overconfident for high predicted MDD probabilities, and fusing these probabilities with the Dirichlet estimate, which its prior shrinks toward $0.5$, moved them closer to the diagonal. Calibration gains therefore come together with the accuracy gains, and they matter for downstream uses of the probability, such as risk communication and deferral thresholds.

\subsection{Explanation Space}
\label{sec:space}

\begin{table}[t]
\centering
\caption{Quality of the retrieved evidence for test subjects with a validation reference bank and $K=5$. Prec.: percentage of retrieved references with the same true label as the query. Acc. and AUC: accuracy and ROC-AUC of $p^{\mathrm{Dir}}_1$ as a stand-alone classifier. Shaded: SHAP space. Bold: best value in each column.}
\label{tab:retrieval}
\setlength{\tabcolsep}{2.2pt}
\resizebox{.9\columnwidth}{!}{%
\begin{tabular}{llrrrrrr}
\toprule
& & \multicolumn{3}{c}{\textbf{TabM}} & \multicolumn{3}{c}{\textbf{ResNet}} \\
\textbf{Space} & \textbf{Metric} & Prec. & Acc. & AUC & Prec. & Acc. & AUC \\
\midrule
Raw features & Cosine & 57.6 & 61.3 & 0.657 & 57.6 & 61.3 & 0.657 \\
Raw features & Pearson & 57.0 & 61.1 & 0.657 & 57.0 & 61.1 & 0.657 \\
Raw features & Euclidean & 56.2 & 60.2 & 0.650 & 56.2 & 60.2 & 0.650 \\
\midrule
\rowcolor{oursbg}SHAP & Cosine & 58.1 & \textbf{62.3} & 0.668 & 58.7 & \textbf{62.5} & \textbf{0.674} \\
\rowcolor{oursbg}SHAP & Pearson & 57.4 & 61.2 & 0.657 & 57.9 & 62.2 & 0.662 \\
\rowcolor{oursbg}SHAP & Euclidean & \textbf{58.9} & 62.2 & \textbf{0.674} & \textbf{58.9} & 61.7 & 0.671 \\
\bottomrule
\end{tabular}}
\end{table}

% \begin{figure}[t]
% \centering
% \includegraphics[width=\linewidth]{img/acm/importance.pdf}
% \caption{The twelve features with the largest mean absolute SHAP value on test subjects, averaged over folds and repetitions. LH/RH: left/right hemisphere. ACC: anterior cingulate cortex. MFG: middle frontal gyrus.}
% \label{fig:importance}
% \end{figure}

Table~\ref{tab:retrieval} compares retrieval in SHAP space with retrieval in the standardized raw feature space. SHAP space with the Euclidean distance gave the most label-consistent references for both backbones (\NTabMPrecShapEuc\% vs.\ at most \NTabMPrecFeatBest\% in feature space), and the Dirichlet estimate reached \NTabMRetShapEuc\% accuracy for TabM, compared with at most \NTabMRetFeatBest\% for any feature-space metric. Explanations remove features that the model ignores and weight the remaining ones by their effect, so neighbors in SHAP space share the evidence the model actually used. A 5-nearest-neighbor classifier on the test subjects of the first TabM repetition reached 63.3\% accuracy in SHAP space and 61.7\% in feature space (Appendix~\ref{app:extra}, Fig.~\ref{fig:embedding}).

The explanations were also consistent across architectures. The rank correlation of the mean absolute SHAP values of TabM and ResNet over the 159 features was \NImpSpearman{}. The most important features (Fig.~\ref{fig:importance}) include the surface area of the left caudal anterior cingulate, the thickness of the rostral anterior cingulate, the caudal middle frontal and precentral cortex, the accumbens volume, and the lingual surface area. The cingulate features overlap with the regions in which large consortia report cortical differences in adult MDD~\cite{Schmaal2017}, and the caudal anterior cingulate, lingual, and precentral features overlap with the variables most often selected by a decision tree for MDD classification in an independent cohort~\cite{Stolicyn2020}. Importance at the level of single folds was less stable, which is expected for small effects estimated from about 270 training subjects.

\subsection{Reliability and Expert Review}
\label{sec:review}

\begin{table}[t]
\centering
\caption{AUROC for separating incorrect from correct baseline decisions on out-of-fold test predictions of all repetitions, with 95\% bootstrap intervals. Shaded: scores that use SHAP-Dirichlet evidence. Bold: best value in each column.}
\label{tab:error}
\small
\setlength{\tabcolsep}{3pt}
\resizebox{\columnwidth}{!}{%
\begin{tabular}{lcc}
\toprule
\textbf{Score} & \textbf{TabM} & \textbf{ResNet} \\
\midrule
Low confidence & 0.648 (0.62--0.67) & 0.641 (0.62--0.67) \\
MC-dropout variance & 0.601 (0.58--0.63) & 0.628 (0.60--0.65) \\
RIS & 0.510 (0.48--0.54) & 0.522 (0.50--0.55) \\
ROS & 0.532 (0.51--0.56) & 0.496 (0.47--0.52) \\
\rowcolor{oursbg}SHAP-Dirichlet (Pearson) & 0.613 (0.59--0.64) & 0.618 (0.59--0.64) \\
\rowcolor{oursbg}SHAP-Dirichlet (Euclidean) & 0.612 (0.59--0.64) & 0.607 (0.58--0.63) \\
\rowcolor{oursbg}Confidence + SHAP-Dirichlet & \textbf{0.659} (0.64--0.68) & \textbf{0.658} (0.63--0.68) \\
\bottomrule
\end{tabular}}
\end{table}

\begin{figure}[t]
\centering
\includegraphics[width=\linewidth]{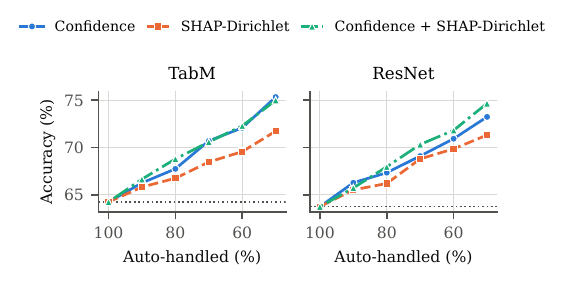}
\caption{Accuracy of the automatically handled test subjects when the least reliable subjects are deferred to expert review, with validation-percentile thresholds. Dotted line: accuracy without review.}
\label{fig:deferral}
\end{figure}

The same signals can decide which subjects need expert review. SHAP-Dirichlet evidence for the opposite class identified baseline errors above chance for both backbones (AUROC \NTabMAucDir{} and \NResNetAucDir{}, Table~\ref{tab:error}), although less well than baseline confidence (\NTabMAucConf{} and \NResNetAucConf). The two signals are complementary. Their combination reached \NTabMAucComb{} and \NResNetAucComb{}, a gain over confidence of $\NResNetAucCombGain$ for ResNet (95\% CI $\NResNetAucCombGainLo$ to $\NResNetAucCombGainHi$) and $\NTabMAucCombGain$ for TabM (95\% CI $\NTabMAucCombGainLo$ to $\NTabMAucCombGainHi$). RIS and ROS alone were close to chance (\NTabMAucRis{}--\NTabMAucRos{} for TabM). They are therefore useful as a gate that limits which decisions may change, but not as error detectors.

Fig.~\ref{fig:deferral} shows the accuracy of the automatically handled subjects when the remaining subjects are deferred. With the combined score, deferring 30\% of the subjects increased this accuracy from \NTabMBase\% to \NTabMDefSeventyComb\% for TabM and from \NResNetBase\% to \NResNetDefSeventyComb\% for ResNet. Deferring half of the subjects gave \NTabMDefHalfComb\% and \NResNetDefHalfComb\%, compared with \NTabMDefHalfConf\% and \NResNetDefHalfConf\% for confidence alone. The SHAP-Dirichlet evidence thus adds an interpretable reason for each referral at no loss in deferral quality.

\begin{figure}[t]
\centering
\includegraphics[width=\linewidth]{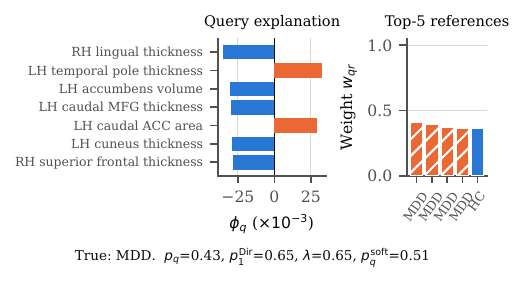}
\caption{A patient with MDD whose TabM baseline prediction was wrong and was corrected by the layer with the enlarged bank. Left: the seven largest SHAP values of the query (orange pushes toward MDD, blue toward HC). Right: weights of the five retrieved references and their true labels.}
\label{fig:case}
\end{figure}

Fig.~\ref{fig:case} shows a corrected case. The baseline predicted HC for a patient with MDD ($p_q=0.43$). The explanation was unstable under small perturbations, and four of its five nearest reference explanations belonged to patients with MDD, so the Dirichlet estimate favored MDD ($p^{\mathrm{Dir}}_1=0.65$). The validation-selected fusion moved the probability to $0.51$, and the corrected decision comes with a set of similar, labeled subjects that a clinician can inspect.

\section{Discussion}
\label{sec:discussion}

\paragraph{What the layer does well} With a reference bank of all labeled subjects, the layer is a safe post-hoc addition to a neural classifier. It changes a few decisions, it acts mostly where the classifier is uncertain, it improves the calibration of the probabilities, and its gains were positive in every neural repetition. The ablations identify the components that make it safe. A gate limits the correction to about a third of the subjects, and soft fusion moves a probability only as far as the evidence supports. Variants without a gate or with hard flips reduced accuracy, because they also change many correct decisions.

\paragraph{Why the reference bank matters} The Dirichlet estimate is an average over $K$ retrieved subjects, so its quality depends on how densely the bank covers explanation space. With about 135 validation subjects, retrieved neighbors are often far from the query, and the evidence is noisy. Adding the training subjects triples the bank. Their explanations come from the same fixed model, and their labels are known, so the enlarged bank uses no test information. The consistent gain in all \NBDeepRepN{} neural repetitions suggests that the layer will benefit from more labeled reference subjects in deployment, for example from other sites.

\paragraph{Why the accuracy gains are modest} The baselines are weak, in line with other structural MRI studies of depression~\cite{Flint2021,Stolicyn2020}, and evidence retrieved from a few hundred subjects cannot be much more accurate than the classifier itself. The Dirichlet estimate alone reached about 62\% accuracy. Each rule is also selected on a small validation set, so the selection is noisy. We therefore present accuracy as one benefit among several, together with calibration, low-confidence correction, and expert triage.

\paragraph{The role of RIS and ROS} RIS and ROS did not separate correct from incorrect decisions on their own, so the gate does not work as an error detector. It works as a selector of subjects whose explanation is sensitive to small, measurement-scale feature changes, which makes the correction conservative. The gated subjects were also less accurate than the rest, so the gate and the evidence act on the same subjects. Monte Carlo dropout variance was an even better gate in our experiments, and the framework accepts any uncertainty score.

\paragraph{Classical classifiers} The layer helped the RBF-SVM but not logistic regression or the random forest. Both of the latter produce probabilities close to $0.5$ for many subjects, and for logistic regression the SHAP vectors are rescaled feature vectors, so explanation retrieval adds little beyond the features themselves. The effect of the layer on a given classifier should therefore be confirmed on independent folds before deployment, as in our protocol.

\paragraph{Lessons for evaluation} Single test folds are too small to rank post-hoc rules on this task. In the first two repetitions, the accuracy change of the twelve fixed rule variants on single folds ranged from $-4.95$ to $+5.0$ points, and the variant with the largest single-fold gain changed accuracy by $-0.59$ points on average over all folds. Selecting a rule by its test-fold gain would therefore have reported a misleading improvement. Repeated nested cross-validation with pre-specified tests and a separate confirmation step is necessary for post-hoc methods on small neuroimaging datasets~\cite{Varoquaux2018,Varoquaux2022}.

\paragraph{Limitations} We used one dataset with site effects that were not harmonized~\cite{FORTIN2018104}, and we did not test transfer to external sites. The enlarged reference bank was chosen after two repetitions and confirmed on two further repetitions of the same dataset, which is an internal rather than an external confirmation. The validation set serves for early stopping and for rule selection, which makes validation estimates optimistic, although the test folds remain untouched. KernelSHAP for perturbed copies is expensive, especially for large ensembles. Finally, the expert-review path was evaluated by simulation and not with clinicians.

\section{Conclusion}
\label{sec:conclusion}

We presented uncertainty-gated SHAP-Dirichlet refinement, a post-hoc layer that retains, refines, or defers the predictions of a fixed MRI-based MDD classifier. In repeated nested cross-validation on 504 SRPBS subjects, the layer with a training-plus-validation reference bank improved accuracy in every TabM and ResNet repetition, concentrated these gains on low-confidence subjects, and improved calibration. Gating and soft fusion were essential for these properties, and the effect carried over to an RBF-SVM but not to logistic regression or a random forest. SHAP-space retrieval produced more label-consistent references than raw-feature retrieval, and the resulting evidence supported a deferral policy that raised the accuracy of automatically handled subjects to about 75\% at 50\% coverage. A natural next step is to validate the review pathway with clinicians, to enlarge the reference bank across sites, and to test the layer under external site and cohort shifts.

\bibliographystyle{ACM-Reference-Format}
\bibliography{ref-acm}

\appendix
% Let the appendix floats share the remaining columns instead of spilling onto an extra page.
\renewcommand{\topfraction}{0.9}\renewcommand{\bottomfraction}{0.9}\renewcommand{\textfraction}{0.05}
\setcounter{topnumber}{3}\setcounter{bottomnumber}{2}\setcounter{totalnumber}{5}
\section{Hyperparameter Grids and Selected Models}
\label{app:grids}

Table~\ref{tab:grids} lists the hyperparameter grids, and Table~\ref{tab:configs} summarizes the models selected by the inner GridSearch. The selected configuration varied across folds for each classifier, and the baseline accuracy of individual outer folds differed by more than 10 percentage points, which is why we report pooled results across all folds and repetitions.

\begin{table}[ht]
\centering
\caption{Hyperparameter grids. All grids were searched with a 3-fold inner cross-validation inside every outer fold, optimizing the F1-score.}
\label{tab:grids}
\small
\setlength{\tabcolsep}{3pt}
\begin{tabular}{>{\raggedright\arraybackslash}p{0.2\linewidth}>{\raggedright\arraybackslash}p{0.74\linewidth}}
\toprule
\textbf{Classifier} & \textbf{Fixed settings and searched values} \\
\midrule
TabM & PLE (20 bins), mini ensemble of 32 members, averaged ensemble outputs, no normalization. Layer sizes [128] or [256], dropout 0.1/0.2/0.3, learning rate $5{\times}10^{-5}$/$10^{-4}$/$5{\times}10^{-4}$/$10^{-3}$, weight decay $10^{-4}$/$5{\times}10^{-4}$/$10^{-3}$ (72 configurations). \\
ResNet & PLE (20 bins), layer normalization, skip connections. Two blocks [128, 64, 32] or three blocks [256, 128, 64], dropout 0.1/0.2/0.3, learning rate as for TabM, weight decay $10^{-5}$/$10^{-4}$/$10^{-3}$ (72 configurations). \\
Logistic regression & Standardized features, L2 penalty, $C\in\{10^{-3},10^{-2},10^{-1},1,10\}$. \\
RBF-SVM & Standardized features, Platt-scaled probabilities, $C\in\{0.1,1,10\}$, $\gamma\in\{\text{scale},10^{-3},10^{-2}\}$. \\
Random forest & 300 trees, maximum depth $\{\text{none},5,10\}$, minimum leaf size $\{1,5\}$, maximum features $\{\sqrt{d},0.3d\}$. \\
\bottomrule
\end{tabular}
\end{table}

\begin{table}[ht]
\centering
\caption{Selected models over all outer folds and repetitions. Distinct: number of distinct selected configurations. Accuracy: baseline test accuracy (\%) of single outer folds.}
\label{tab:configs}
\small
\setlength{\tabcolsep}{3pt}
\resizebox{.9\columnwidth}{!}{%
\begin{tabular}{lrrcc}
\toprule
\textbf{Classifier} & \textbf{Folds} & \textbf{Distinct} & \textbf{Mean $\pm$ SD} & \textbf{Range} \\
\midrule
TabM & 20 & 18 & 64.24 $\pm$ 4.48 & 56.00--73.27 \\
ResNet & 20 & 18 & 63.74 $\pm$ 3.42 & 56.00--69.00 \\
Logistic regression & 10 & 3 & 64.28 $\pm$ 3.89 & 58.00--70.30 \\
RBF-SVM & 10 & 6 & 65.87 $\pm$ 2.44 & 61.39--69.31 \\
Random forest & 10 & 6 & 66.28 $\pm$ 3.08 & 61.39--70.00 \\
\bottomrule
\end{tabular}}
\end{table}

\section{Additional Results}
\label{app:extra}

\begin{figure}[t]
\centering
\includegraphics[width=.7\linewidth]{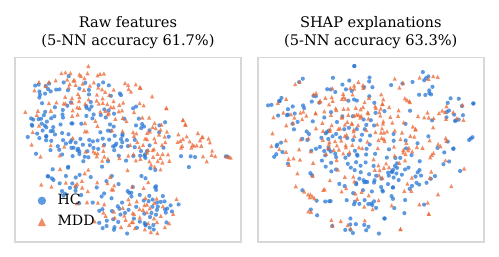}
\caption{t-SNE of the 504 test subjects (first TabM repetition) in feature space and SHAP space, with the 5-fold cross-validated 5-NN accuracy of each space.}
\label{fig:embedding}
\end{figure}

Table~\ref{tab:reps} reports every repetition for both reference banks, Table~\ref{tab:ablation_classic} repeats the ablations for the classical classifiers, and Fig.~\ref{fig:embedding} compares feature space and SHAP space with t-SNE.

\begin{table}[ht]
\centering
\caption{Per-repetition results for the validation (Val.) and training+validation (Train+val.) banks. Base: pooled baseline accuracy (\%). $\Delta$: accuracy change (pp), \gain{green} for a gain and \loss{red} for a loss. C/B: corrected/broken decisions. $p$: exact McNemar test.}
\label{tab:reps}
\small
\setlength{\tabcolsep}{2.4pt}
\resizebox{.8\columnwidth}{!}{%
\begin{tabular}{lcrrrrrrr}
\toprule
& & & \multicolumn{3}{c}{\textbf{Val.\ bank}} & \multicolumn{3}{c}{\textbf{Train+val.\ bank}} \\
\textbf{Model} & \textbf{Seed} & \textbf{Base} & $\Delta$ & C/B & $p$ & $\Delta$ & C/B & $p$ \\
\midrule
TabM & 42 & 65.87 & \gain{+0.20} & 7/6 & 1.00 & \gain{+0.60} & 4/1 & 0.38 \\
TabM & 43 & 63.69 & \gain{+1.59} & 11/3 & 0.06 & \gain{+0.60} & 11/8 & 0.65 \\
TabM & 44 & 62.70 & \loss{$-$0.40} & 10/12 & 0.83 & \gain{+1.98} & 26/16 & 0.16 \\
TabM & 45 & 64.68 & \loss{$-$1.19} & 5/11 & 0.21 & \gain{+0.99} & 13/8 & 0.38 \\
\midrule
ResNet & 42 & 63.10 & \gain{+0.40} & 14/12 & 0.85 & \gain{+0.40} & 19/17 & 0.87 \\
ResNet & 43 & 64.68 & \gain{+0.60} & 14/11 & 0.69 & \gain{+0.79} & 14/10 & 0.54 \\
ResNet & 44 & 62.50 & \gain{+1.19} & 16/10 & 0.33 & \gain{+0.40} & 6/4 & 0.75 \\
ResNet & 45 & 64.68 & \loss{$-$0.20} & 9/10 & 1.00 & \gain{+0.20} & 6/5 & 1.00 \\
\midrule
LogReg & 42 & 64.88 & \loss{$-$0.99} & 7/12 & 0.36 & \gain{+0.40} & 12/10 & 0.83 \\
LogReg & 43 & 63.69 & \gain{+0.20} & 13/12 & 1.00 & \loss{$-$0.60} & 10/13 & 0.68 \\
\midrule
SVM & 42 & 65.48 & +0.00 & 14/14 & 1.00 & \gain{+1.39} & 22/15 & 0.32 \\
SVM & 43 & 66.27 & \gain{+2.18} & 19/8 & 0.05 & \gain{+1.39} & 14/7 & 0.19 \\
\midrule
RF & 42 & 66.67 & \loss{$-$0.99} & 11/16 & 0.44 & \loss{$-$0.20} & 6/7 & 1.00 \\
RF & 43 & 65.87 & \loss{$-$0.99} & 7/12 & 0.36 & \loss{$-$0.79} & 4/8 & 0.39 \\
\bottomrule
\end{tabular}}
\end{table}

\begin{table}[!tb]
\centering
\caption{Ablations for the classical classifiers (two repetitions). $\Delta$: mean accuracy change (pp), \gain{green} for a gain and \loss{red} for a loss. I/R: repetitions with improved accuracy. Shaded: the pre-specified layer.}
\label{tab:ablation_classic}
\small
\setlength{\tabcolsep}{1.6pt}
\resizebox{\columnwidth}{!}{%
\begin{tabular}{llrrrrrr}
\toprule
& & \multicolumn{2}{c}{\textbf{LogReg}} & \multicolumn{2}{c}{\textbf{SVM}} & \multicolumn{2}{c}{\textbf{RF}} \\
\textbf{Comp.} & \textbf{Variant} & $\Delta$ & I/R & $\Delta$ & I/R & $\Delta$ & I/R \\
\midrule
\multirow{3}{*}{Correction} & Hard flip & \loss{$-$0.10} & 0/2 & \gain{+0.30} & 1/2 & \loss{$-$0.99} & 0/2 \\
 & Hard or soft (validated) & \loss{$-$0.10} & 1/2 & \gain{+0.99} & 1/2 & \loss{$-$1.29} & 0/2 \\
 & \ours Soft fusion (ours) & \ours \loss{$-$0.40} & \ours 1/2 & \ours \gain{+1.09} & \ours 1/2 & \ours \loss{$-$0.99} & \ours 0/2 \\
\midrule
\multirow{6}{*}{Gate} & None & \loss{$-$0.89} & 1/2 & \gain{+0.40} & 1/2 & \loss{$-$1.29} & 0/2 \\
 & Random & \gain{+0.02} & 1/2 & \gain{+0.05} & 1/2 & \loss{$-$0.90} & 0/2 \\
 & Low confidence & \loss{$-$0.10} & 1/2 & \gain{+1.09} & 2/2 & \gain{+0.20} & 1/2 \\
 & RIS only & \loss{$-$0.30} & 0/2 & \gain{+0.69} & 2/2 & \loss{$-$0.99} & 0/2 \\
 & ROS only & \loss{$-$0.10} & 1/2 & \gain{+0.60} & 2/2 & \loss{$-$0.99} & 0/2 \\
 & \ours RIS/ROS, validated (ours) & \ours \loss{$-$0.40} & \ours 1/2 & \ours \gain{+1.09} & \ours 1/2 & \ours \loss{$-$0.99} & \ours 0/2 \\
\midrule
\multirow{2}{*}{Evidence} & Raw-feature retrieval & \gain{+0.79} & 2/2 & \gain{+0.20} & 1/2 & \loss{$-$0.20} & 1/2 \\
 & \ours SHAP, validation bank (ours) & \ours \loss{$-$0.40} & \ours 1/2 & \ours \gain{+1.09} & \ours 1/2 & \ours \loss{$-$0.99} & \ours 0/2 \\
\bottomrule
\end{tabular}}
\end{table}

\clearpage

\end{document}